\documentclass[graybox]{svmult}

\usepackage[utf8]{inputenc}

\usepackage{mathptmx}       
\usepackage{helvet}         
\usepackage{courier}        
\usepackage{type1cm}        
\usepackage{makeidx}         
\usepackage{graphicx}        
\usepackage{multicol}        
\usepackage[bottom]{footmisc}

\usepackage{listings}
\usepackage{dsfont}
\usepackage[numbers]{natbib}

\makeindex             

\begin{document}

\title*{A quantum mechanical bound for CHSH-type Bell inequalities}
\author{Michael Epping, Hermann Kampermann and Dagmar Bru\ss{}}
\institute{Michael Epping \and Hermann Kampermann \and Dagmar Bru\ss{} \at Heinrich-Heine-University D\"usseldorf, Universit\"atsstr. 1, 40225 D\"usseldorf, Germany, \email{epping@hhu.de}}
%
%
\maketitle

\abstract{Many typical Bell experiments can be described as follows. A source repeatedly distributes particles among two spacelike separated
observers. Each of them makes a measurement, using an observable randomly chosen out of several possible ones, leading to one of two
possible outcomes. After collecting a sufficient amount of data one calculates the value of a so-called Bell expression. An important
question in this context is whether the result is compatible with
bounds based on the assumptions of locality, realism and freedom of choice. Here we are interested in bounds on the obtained value derived
from quantum theory, so-called Tsirelson bounds. We describe a simple Tsirelson bound, which is based on a singular value decomposition.
This mathematical result leads to some physical insights. In particular the optimal observables can be obtained. Furthermore statements
about the dimension of the underlying Hilbert space are possible. Finally, Bell inequalities can be modified to match rotated measurement
settings, e.g. if the two parties do not share a
common reference frame.}

\section{Introduction}
Since the advent of quantum theory physicists have been struggling for a deeper understanding of its concepts and implications. One approach
to this end is to carve out the differences between quantum theory and ``classical'' theories, i.e. to explicitly point to the conflicts
between quantum theory and popular preconceptions, which evolved in each individual and the scientific community from decoherent macroscopic
experiences. Plain formulations of such discrepancies and convincing experimental demonstrations are crucial to internalizing quantum
theory and replacing existing misconceptions. For this reason the double-slit-experiments (and similar experiments with optical gratings) 
\cite{YoungLectures,FeynmanLectures3,NJP153033018,PhysRevLett.88.100404,NaturePhysics.10.271}, which expose the role
of state superpositions in quantum theory, are so very fascinating and famous. Other examples of ``eye-openers'' are demonstrations of
tunneling~\citetext{\citealp{Razavy}, \citealp[pp. 33-12]{FeynmanLectures2}}, the quantum Zeno effect~\cite{PhysRevA.41.2295} and variations
of the Elitzur-Vaidman-scheme~\cite{BombTester,PhysRevLett.83.4725,QuantumImaging}, to
pick just a few.\\ 
Bell experiments~\cite{PhysRevLett.49.91,PhysRevLett.81.5039,Wineland,Nature461.504}, which show entanglement in a particularly striking
way, belong to
this list. Informally, entanglement is the fact
that in quantum theory the state of a compound system (e.g. two particles) is not only a collection of the states of the subsystems. This fact can lead to strong correlations between measurements on different subsystems. Before going into more
detail here, we would like to note that the described differences between the relatively new quantum theory and our old preconceptions are
obvious starting points when to look for innovative technologies which were even unthinkable before. This is in fact a huge motivation for
the field of quantum information, where Bell experiments play a central role.\\
\subsection{Bell experiments bring three fundamental common sense assumptions to a test}
\begin{figure}[tbp]%
\centering
\includegraphics[scale=0.25]{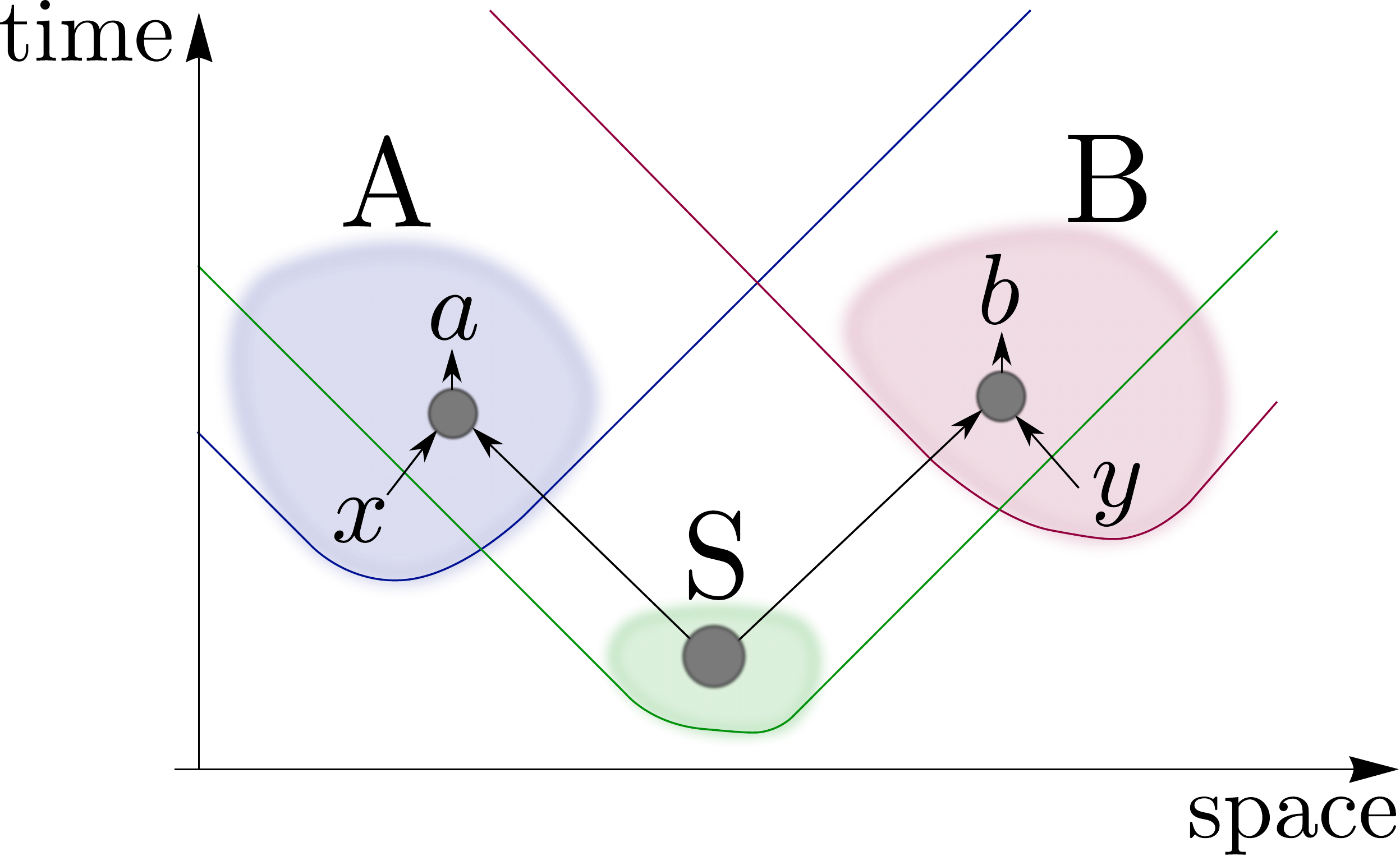} %
\caption{Two parties, Alice (A) and Bob (B), perform a Bell experiment. Both of them receive parts of a
quantum system from the source (S).
They randomly choose a measurement setting, denoted by $x=1,2,...,M_1$ and $y=1,2,...,M_2$, and write down their outcomes $a=-1 \mbox{ or
}1$ and $b=-1\mbox{ or }1$, respectively. The experiment is repeated until the accumulated data is analyzed according to the text. Angles of
45 degrees in the space-time-diagram correspond to the speed of light. The future light cones of A, B and S show, that the setting choice
and outcome of one party
cannot influence the other and that A and B also cannot influence any event inside the source.}\label{fig:setup}%
\end{figure}%
The idea of Bell was to show that some common sense assumptions lead to predictions of experimental data which contradict the predictions
of quantum theory. 
In the following we employ a black box approach to emphasize that this idea is completely independent of the physical realization of
an experiment. For example the measurement apparatuses get some input (an integer number which will in the following be called ``setting'')
and produce some output (the ``measurement outcomes''). We refer readers
preferring a more concrete notion to Section~\ref{sec:CHSHexperimentell}, where physical implementations and
concrete measurements are outlined.\\
In the present paper we consider the following (typical) Bell experiment, see also Figure~\ref{fig:setup}. There are
three experimental sites, two of which we call the parties Alice (A) and Bob (B), and the third being a preparation site which we call
source (S). Alice and Bob have a spatial separation large enough such that no signal can travel from one party to the other at the speed of
light during the execution of our experiment. The source is separated such that no signal can travel from A or B to it at the speed of light
before it finishes the state production. The importance of such separations will become clear later.\\
The source produces a quantum system, and sends one part to Alice and one to Bob. We will exemplify this in Section~\ref{sec:CHSHexperimentell}. A and B are in possession of measurement apparatuses with a predefined set of different settings. In each run they choose the setting randomly, e.g. they turn a knob located at the outside of the
apparatus, measure the system received from the source and list the setting and outcome. In the present paper the
measurements are two-valued and the outcomes are denoted by $-1$ and $+1$. Let  $M_1$ and $M_2$ be the number of different measurement
settings at site A and B, respectively. We label them by $x=1,2,3,...,M_1$ for Alice and $y=1,2,3,...,M_2$ for Bob.
This preparation and measurement procedure of a quantum system is repeated until the amount of data suffices to estimate the expectation
value of the measured observables, up to the statistical accuracy one aims at. The expectation value of an observable is the average of all
possible outcomes, here $\pm 1$, weighted with the corresponding probability to get this outcome. \\
Let us sketch the preconceptions that are \textit{jointly} in conflict with the quantum theoretical predictions for Bell tests.
These are mainly three concepts: Locality, realism and freedom of choice. This forces us to question at least one of these ideas, because
any interpretation of quantum theory, as well as any ``postquantum'' theory, cannot obey all of them. We invite the reader to pick one to
abandon while reading the following descriptions. Do not be confused by our comparison with the textbook formalism of quantum theory: so far
you are free to choose any of them.\\
{\em Locality} is the assumption, that effects only have nearby direct causes, or the other way around: any action can only affect directly
nearby objects. If some action here has an impact there, then something traveled from here to there. And, according to special relativity,
the
speed of this signal is at most the speed of light. In our setup, this means that whatever Alice does cannot have any observable effect at
Bob's site. In particular, the measurement outcome at one side cannot depend on the choice of measurement setting at the other site. While
the formalism of quantum theory has some ``nonlocal features'', e.g. a global state, it is strictly local in the above sense, because any
local quantum operation on one subsystem does not change expectation values of local observables for a different subsystem.\\
{\em Realism} is the concept of an objective world that exists independently of subjects (``observers''). A stronger form of realism is the
``value-definiteness'' assumption meaning that the properties of objects always have definite values, also if they are not measured or even
unaccessible for any observer. It seems to be against common sense to assume that objects cease to have definite properties if we do not
measure them any longer. In particular the natural sciences were founded on the assumption, that nature and its properties exist
independently of the scientist. In our setup realism implies, that the measurement outcomes of unperformed measurements (in unchosen
settings) have some value. We do not know them, but we can safely assume that they exist, give them a name and use them as variables. If
possible outcomes are $-1$ and $+1$, for example, we might use that the outcome squared is $1$ in any of our calculations. In general the
(usual) formalism of quantum theory does not contain definite values for measurement outcomes independent of a measurement. \\
{\em Freedom of choice}, which is also sometimes called the free will assumption, means it is possible to freely choose what experiment to
perform
and how. Because this idea is elusive, we are content with a decision that is statistically independent of any quantity which is subject
of our experiment. The idea of fate seems to be tempting to many people. However, dropping freedom of choice makes science useless. Just
imagine you ``want'' to investigate the question whether a bag contains black balls but your fate
is to pick only white balls (and put them back afterwards), even though there are many black balls inside. In our setup, freedom of choice
implies, that A's and B's choice of measurement setting does not depend on the other's choice or the outcomes. In quantum theory, there is
freedom of choice in the sense that random measurement outcomes of some other process can be used to make decisions.\\
If you decided that you preferably take leave of locality you are in good company. Many scientists conclude from Bell's theorem, that
the locality assumption is not sustainable. This is particularly interesting when you consider the above comparison with the standard
textbook formalism of quantum theory, which is apparently not realistic but local in the described sense. The fact that in this context many
scientists speak about ``quantum nonlocality'' thus leads to controversy~\cite{JPA46424009}. We therefore want to stress again, that the
experimental contradiction only tells us that at least one of all the assumptions that lead to the predictions needs to be wrong. We
cannot decide which assumption is wrong from Bell's theorem alone.\\
We now focus on a tool to show the contradiction in the described experiment between the above assumptions and quantum theory, the so called
Bell inequalities. These are inequalities of measurable quantities which are (mainly) derived from locality, realism and freedom of choice
and therefore hold for all theories which obey these principles, while they are violated by the predictions of quantum theory. We consider a
special kind of Bell inequalities which are linear combinations of joint expectation values of Alice's and Bob's observables. The joint
expectation value of the two observables of Alice and Bob is the expectation value of the product of the measurement
outcomes, which again takes values $\pm 1$. It depends on the setting choice $x$ at Alice's site and $y$ at Bob's site and we denote it by
$E(x,y)$. If we denote
the (real) coefficient in front of the expectation value $E(x,y)$ as $g_{x,y}$, then we
can write such Bell inequalities as
\begin{equation}
 \sum_{x=1}^{M_1}\sum_{y=1}^{M_2} g_{x,y} E(x,y) \leq B_{g}, \label{eq:bellineq}
\end{equation}
where the bound $B_g$ depends only on the coefficients $g_{x,y}$. These coefficients form a matrix $g$ which has dimension $M_1\times
M_2$. Any real matrix $g$ defines a Bell inequality via Eq.~(\ref{eq:bellineq}). The may be most famous example is the
Clauser-Horne-Shimony-Holt (CHSH) \cite{PhysRevLett.23.880} inequality, which reads
\begin{equation}
 E(1,1)+E(1,2)+E(2,1)-E(2,2) \leq 2. \label{eq:CHSH}
\end{equation}
Here the corresponding matrix $g$ is
\begin{equation}
 g=\left(\begin{array}{cc}
          1 & 1\\ 1 & -1
         \end{array}\right). \label{eq:gCHSH}
\end{equation}
Due to its prominence we call the class of Bell inequalities in the form of Eq.~(\ref{eq:bellineq}) CHSH-type Bell inequalities. For
completeness we sketch the derivation of $B_g$. It turns out that it suffices to consider deterministic outcomes only,
as a probabilistic theory, where the outcomes follow some probability distribution, cannot achieve a higher value in Eq.~(\ref{eq:bellineq}): it can be described as a mixture of deterministic
theories and the value of Eq.~(\ref{eq:bellineq}) is the sum of the values for the deterministic theories weighted with the corresponding
probability in the mixture. For deterministic theories the expectation value is merely the product of the two (possibly unmeasured)
outcomes $a$ of Alice and $b$ of Bob, which we are allowed to use when assuming {\em realism}. Due to {\em locality} $a$ only depends on the
setting $x$ of Alice, which has no further dependence due to {\em freedom of choice}. Analogously $b$ depends only on the setting $y$ of
Bob, which in turn
has no further dependence. Thus the expectation value is 
\begin{equation}
 E(x,y)=a(x) b(y). \label{eq:Eclass}
\end{equation}
Now we can calculate $B_g$ by maximizing
Eq.~(\ref{eq:bellineq}) over all possible assignments of $-1$ and $+1$ values to $a(x)$ and $b(y)$. In Eq.~(\ref{eq:CHSH}) the maximal value
is $B_g=2$, which is achieved for $a(1)=a(2)=b(1)=b(2)=1$, for example. Note that the sign of $E(2,2)$ cannot be changed independently of the other three terms, because $E(1,2)$ and $E(2,1)$ contain $b(2)$ and $a(2)$, respectively.\\
We point out that any function that maps the probabilities of different measurement outcomes to a real number may be
used to derive Bell inequalities, and different types of Bell inequalities can be found in the literature (e.g.~\cite{PhysRevD.10.526}).
However, here we focus on Bell inequalities of the form of Eq.~(\ref{eq:bellineq}).\\
\subsection{The CHSH inequality can be violated in experiments with entangled photons}\label{sec:CHSHexperimentell}
We recapitulate some basics of quantum (information) theory. Analogously to a classical bit the quantum bit, or qubit, can be in two states
$0$ and $1$, but additionally in every possible superposition of them. Mathematically this state is a unit vector in the two-dimensional
Hilbert space (a vector space with a scalar product) $\mathds{C}^2$ spanned by the basis vectors
\begin{equation}
 \vec{0}:=\left(\begin{array}{c}
                 1 \\ 0
                \end{array}\right) \;\mbox{ and }\; \vec{1}:=\left(\begin{array}{c}
                 0 \\ 1
                \end{array}\right).
\end{equation}
An example of a superposition of these basis states is $\vec{\psi}=\frac{1}{\sqrt{2}}(\vec{0}+\vec{1})$. Any observable on a qubit with
outcomes $+1$
and $-1$ can be written as
\begin{equation}
 A= a_x \underbrace{\left(\begin{array}{cc}
                         0 & 1\\ 1 & 0
                        \end{array}\right)}_{\sigma_x}+a_y \underbrace{\left(\begin{array}{cc}
                         0 & -i\\ i & 0
                        \end{array}\right)}_{\sigma_y}+a_z \underbrace{\left(\begin{array}{cc}
                         1 & 0\\ 0 & -1
                        \end{array}\right)}_{\sigma_z}, \label{eq:obs}
\end{equation}
where the vector $\vec{a}=(a_x,a_y,a_z)^T$ (here $^T$ denotes transposition) defines the measurement direction and the matrices $\sigma_x$,
$\sigma_y$ and $\sigma_z$ are
called Pauli matrices. The expectation value of this observable given any state $\vec{\psi}$ can be calculated as $E=\vec{\psi}^\dagger
A \vec{\psi}$ (here $^\dagger$ denotes the complex conjugated transpose), which is between $-1$ and $+1$.\\
Any quantum mechanical system with (at least) two degrees of freedom can be used as a qubit. In the present context the spin of a
spin-$\frac{1}{2}$-particle, two energy levels of an atom and the polarization of a photon are important examples of qubits. The spin
measurement can be performed using a Stern-Gerlach-Apparatus~\cite{GerthsenPhysik}, the energy level of an atom may be measured using resonant laser
light, or the polarization of a photon can be measured using polarization filters or polarizing beam
splitters.\\
The Hilbert space of two qubits is constructed using the tensor product, i.e. $\mathds{C}^2\otimes \mathds{C}^2=\mathds{C}^4$. The tensor
product
of two matrices (of which vectors are a special case) is formed by multiplying each component of the first matrix with the complete second
matrix, such that a bigger matrix arises. The state of the composite system of two qubits in states $\vec{\phi}^A=(\phi^A_1,\phi^A_2)^T$ and
$\vec{\phi}^B=(\phi^B_1,\phi^B_2)^T$ then reads $\vec{\phi}^{AB}=\vec{\phi}^A\otimes \vec{\phi}^B=(\phi^A_1 \phi^B_1,\phi^A_1
\phi^B_2,\phi^A_2 \phi^B_1,\phi^A_2 \phi^B_2)^T$. The states of such composite systems might be superposed, which leads to the notion of entanglement.\\
Out of several physical implementations of the CHSH experiment we sketch the ones with polarization entangled photons (see
\cite{UndergraduateLaboratory}). We identify $\vec{0}$ with the horizontal and $\vec{1}$ with the vertical polarization of a photon.
Nonlinear processes in special optical elements can be used to create two photons in the state 
\begin{equation}
\vec{\phi}_+ =\frac{1}{\sqrt{2}} (1,0,0,1)^T,
\end{equation}
i.e. an equal superposition of two horizontally polarized photons and two vertically polarized photons. The measurements of Alice and Bob in
setting $1$ and $2$ are
\begin{eqnarray}
 A_1 &=& \cos(2\times 22.5^\circ) \sigma_x + \sin(2\times 22.5^\circ)\sigma_z,\\
 A_2 &=& \cos(-2\times 22.5^\circ) \sigma_x + \sin(-2\times 22.5^\circ)\sigma_z,\\
 B_1 &=& \cos(2\times 0^\circ) \sigma_x + \sin(2\times 0^\circ)\sigma_z\\
 \mbox{and }B_2 &=& \cos(2\times 45^\circ) \sigma_x + \sin(2\times 45^\circ)\sigma_z,
\end{eqnarray}
respectively. Here the angles are the angles of the polarizer and the factor $2$ is due to the fact that in contrast to the
Stern-Gerlach-Apparatus a rotation of the polarizer of $180^\circ$ corresponds to the same measurement again. One can now calculate the
value of Eq.~\ref{eq:CHSH}:
\begin{eqnarray}
 E(1,1)+E(1,2)+E(2,1)-E(2,2)&=&\phantom{+{}}\vec{\phi}_+^\dagger (A_1\otimes B_1) \vec{\phi}_++\vec{\phi}_+^\dagger
(A_1\otimes
B_2) \vec{\phi}_+\nonumber \\
& &+\vec{\phi}_+^\dagger (A_2\otimes B_1) \vec{\phi}_+-\vec{\phi}_+^\dagger (A_2\otimes
B_2) \vec{\phi}_+\nonumber\\
&=&\frac{1}{\sqrt{2}}+\frac{1}{\sqrt{2}}+\frac{1}{\sqrt{2}}-\left(-\frac{1}{\sqrt{2}}\right)\nonumber\\
&=&2\sqrt{2}.
\end{eqnarray}
The value $2\sqrt{2}\approx 2.82$ is larger than $2$ and therefore the CHSH inequality is violated. One can ask whether it is possible to
achieve an even higher value, e.g. when using higher-dimensional systems than qubits, because at the first glance a value of up to four
seems to be possible. This question is addressed in the following sections (the answer, which is negative, is given in
Section~\ref{sec:svbound}).
\subsection{The quantum analog to classical bounds on Bell Inequalities are Tsirelson Bounds}
Analogously to the ``classical'' bound one can ask for bounds on the maximal value of a Bell inequality obtainable within quantum
theory, so-called Tsirelson bounds~\cite{Cirel'son1980}, and the observables that should be measured to achieve this value. In other words:
which
observables are best suited to show the contradiction between quantum theory and the conjunction of the three discussed common sense
assumptions. This question, which is also of some importance for applications of Bell inequalities, is the main subject of the present
essay.\\
The scientific literature contains several approaches to derive Tsirelson bounds, some of which we want to mention. The problem of finding
the Tsirelson bound of Eq.~(\ref{eq:bellineq}) can be formulated as a semidefinite program. Semidefinite programming is a method to obtain
the global optimum of functions, under the restriction that the variable is a positive semidefinite matrix (i.e. it has no negative
eigenvalues). This implies that well developed (mostly numerical) methods can be applied~\cite{Wehner,PhysRevLett.98.010401}. The interested
reader can find a Matlab code snippet to play around with in the appendix~\ref{sec:snippets}. Furthermore there has been some effort to
derive Tsirelson bounds
from
first principles, amongst them the non-signalling principle~\cite{PopescuRohrlich}, information
causality~\cite{InformationCausality} and the exclusivity principle~\cite{PhysRevLett.110.060402}.\\ %
The non-signalling principle is satisfied by all theories, that do not allow for faster-than-light communication. Information causality is
a generalization of the non-signalling principle, in which the amount of information one party can gain about data of another is restricted
by the amount of (classical) communication between them. The exclusivity principle states, that the probability to see one event out of a
set of pairwise exclusive events cannot be larger than one.
\section{The singular value bound}
Here we will discuss a simple mathematical bound for the maximal quantum value of a CHSH-type Bell inequality defined via a matrix
$g$, which we derived in \cite{PhysRevLett.111.240404}. While
it is not as widely applicable as the semidefinite
programming approach, it is an analytical expression which is easy to calculate and it already enables valuable insights. For ``simple'' Bell inequalities, like the CHSH inequality given above, it
is sufficient to use the method of this paper.\\
We will make use of singular value decompositions of real matrices, a standard tool of linear algebra, which we now shortly recapitulate.
\subsection{Any matrix can be written in a singular value decomposition}
A singular value decomposition is very similar to an eigenvalue decomposition, in fact the two concepts are strongly related. Any real
matrix $g$ of dimension $M_1\times M_2$ can be written as the product of three matrices $V$, $S$, $W^T$, i.e. 
\begin{equation}
g=V S W^T, \label{eq:SVDofg}
\end{equation}
where these three matrices have special properties. The matrix $V$ is orthogonal,
i.e. its columns, which are called left singular vectors, are orthonormal. It has a dimension of $M_1\times M_1$. The matrix $S$ is a
diagonal matrix of dimension $M_1\times M_2$, which is not necessarily a square matrix. Its diagonal entries are positive and have
non-increasing order (from upper left to lower right). They are called singular values of $g$. The matrix $W$ is again orthogonal. It has
dimension $M_2\times M_2$ and its columns are called right singular vectors.\\
\begin{figure}
 \begin{center}
 \includegraphics[scale=0.6]{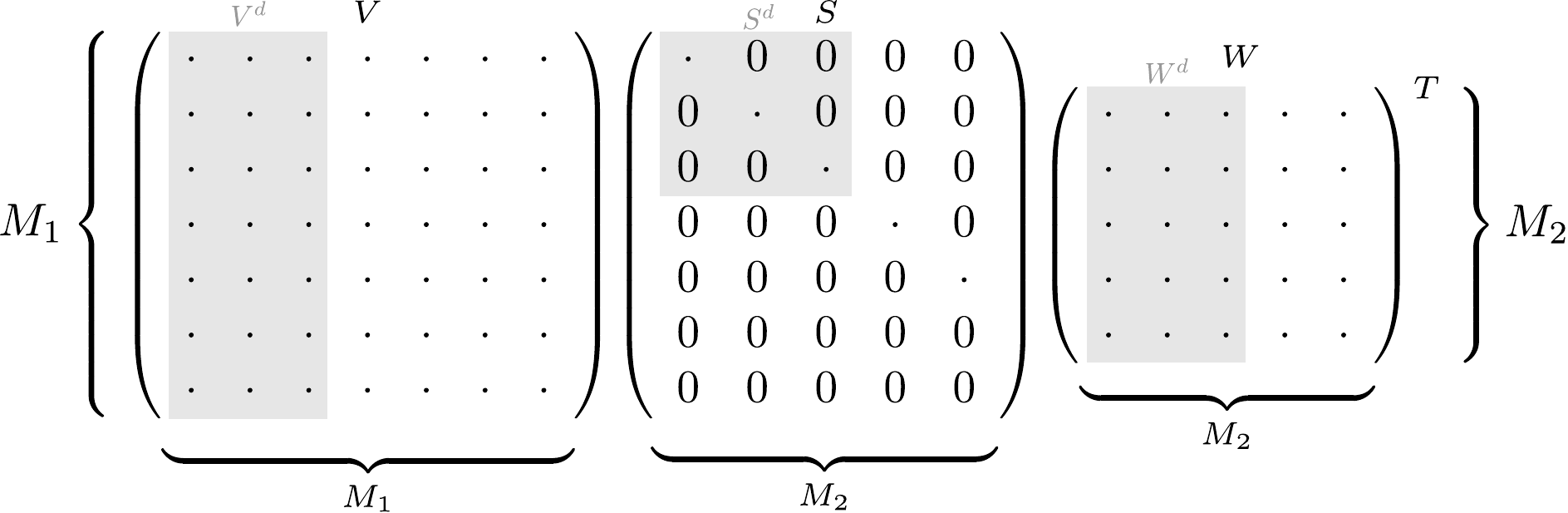}
 \end{center}
 \caption{The matrices involved in the singular value decomposition of a general real $M_1\times M_2$ matrix $g$: $V$ and $W$ are
orthogonal matrices, $S$ is diagonal. $V$ and $W$ contain the left and right singular vectors, respectively, as columns, and $S$ contains
the singular values on its diagonal. The shaded parts belong to a truncated singular value decomposition of $g$. We denote the parts
corresponding to the maximal singular value as $V^{(d)}$, $S^{(d)}$ and $W^{(d)}$.}\label{fig:matrices}
\end{figure}
The largest singular value can appear several times on the diagonal of $S$. We call the number of appearances the degeneracy $d$ of the
maximal singular value. Due to the ordering of $S$, these are the first $d$ diagonal elements of $S$. Here we note the concept of a truncated
singular value decomposition: instead of using the full decomposition one can approximate $g$ by using only parts of the matrices
corresponding to, e.g., the first $d$ singular values (i.e. only the maximal ones). These are the first $d$ left and right singular vectors,
and the first part of $S$, which is just a $d\times d$ identity matrix multiplied by the largest singular value. Since these matrices
play an important role in the following analysis we will give them special names: $V^{(d)}$, $S^{(d)}$ and $W^{(d)}$. All these matrices are
depicted
in Figure~\ref{fig:matrices}.\\
The matrix $g$ maps a vector $\vec{v}$ to a vector $g\vec{v}$ which, in general, has a different length than $\vec{v}$. Here the length is measured by the (usual) Euclidean
norm $||\vec{v}||_2=\sqrt{v_1^2+v_2^2+...+v_{M_2}^2}$. The largest possible stretching factor for all vectors $\vec{v}$ is
a property of the matrix: its matrix norm induced by the Euclidean norm. The value of this matrix norm coincides with the maximal singular
value $S_{11}$. We can therefore express the maximal singular value using
\begin{equation}
 S_{11}=\max_{\vec{v}\in\mathds{R}^{M_2}}  \frac{||g \vec{v}||_2}{||\vec{v}||_2} =: ||g||_2.
\end{equation}
The notation $||g||_2$ for the maximal singular value of $g$ is more convenient than $S_{11}$, as it contains the matrix as an
argument.\\
\subsection{The singular value bound is a simple Tsirelson bound}\label{sec:svbound}
It turns out that the matrix norm of $g$, i.e. its maximal singular value, leads to an upper bound on the quantum value for a Bell
inequality, defined by $g$ via Eq.~(\ref{eq:bellineq}). This is the central insight of this essay. It is remarkable that a {\em mathematical}
property, solely due to the rules of linear algebra, leads to a bound for a {\em physical} theory, here the theory of quantum mechanics.
With the definition of the matrix norm given above, we can now write this singular value bound of $g$, a simple Tsirelson bound~\cite{PhysRevLett.111.240404}. It reads
\begin{equation}
 \sum_{x=1}^{M_1} \sum_{y=1}^{M_2} g_{x_1,x_2} E(x_1,x_2) \leq \sqrt{M_1 M_2} ||g||_2, \label{eq:SVbound}
\end{equation}
where $E$ now denotes the expectation value of a quantum measurement in setting $x_1$ and $x_2$. Eq.~(\ref{eq:SVbound}) is the central
formula of this essay. Note that this bound is not always tight, i.e. there exist examples where the right hand side cannot be
reached within quantum mechanics. However for many examples it is tight. The proof of this bound is sketched in
Appendix~\ref{sec:proofofbound}.\\
We now calculate this bound for the CHSH inequality given in Eq.~(\ref{eq:CHSH}). We
see, that here
the matrix of coefficients is
\begin{equation}
 g=\left(\begin{array}{cc}
          1 & 1\\ 1 & -1
         \end{array}\right) = \underbrace{\left(\begin{array}{cc}
          \frac{1}{\sqrt{2}} & \frac{1}{\sqrt{2}}\\ \frac{1}{\sqrt{2}} & -\frac{1}{\sqrt{2}}
         \end{array}\right)}_{V} \underbrace{\left(\begin{array}{cc} 
\sqrt{2} & 0 \\
0 & \sqrt{2}
\end{array}\right)}_{S} \underbrace{\left(\begin{array}{cc} 
1 & 0 \\
0 & 1
\end{array}\right)}_{W^T}. \label{eq:SVDCHSH}
\end{equation}
It is easy to check that the given decomposition of $g$ is a singular value decomposition, i.e. $V$, $S$ and $W$ have the properties
described above. From this we read, that the maximal singular value of $g$ is $||g||_2=\sqrt{2}$. Then Eq.~(\ref{eq:SVbound}) tells us, that the
maximal value of the CHSH inequality (Eq.~(\ref{eq:CHSH})) within quantum theory is not larger than $2 \sqrt{2}$, a value which can
also be achieved when using
appropriate measurements and states (see Section~\ref{sec:CHSHexperimentell})\\

\subsection{Tightness of the bound can be checked efficiently}
We already mentioned that the inequality~(\ref{eq:SVbound}) is not always tight, i.e. sometimes it is not possible to find observables and
a quantum state such that there is equality. From the derivation of Eq.~(\ref{eq:SVbound}) sketched in Appendix~\ref{sec:proofofbound} one
understands, why this is the case. The value $\sqrt{M_1 M_2}||g||_2$ is achieved if and only if there exists a right singular vector
$\vec{v}$ to the maximal singular value and and a corresponding left singular vector $\vec{w}$ which fulfill further normalization
constraints.\\
It is common to denote the element in the $i$-th row and $j$-th column of a matrix $A$ as $A_{ij}$. We will extend this notation
to denote the whole $i$-th row by $A_{i*}$ and the whole $j$-th column by $A_{*j}$, i.e. the $*$ stands for ``all''. For example, the
$l$-th $M_1+M_2$ dimensional canonical basis vector, with a one at position $l$ and $0$ everywhere else, can then be written as
$\mathds{1}^{(M_1+M_2)}_{*,l}$.\\
With this notation at hand we write down the normalization constraint from above as the system of equations
\begin{eqnarray}
 \left\|\alpha^T V_{x*}^{(d)} \right\|^2 =& 1  \mbox{ for }x=1,2,...,M_1 \label{eq:normv}\\
\mbox{and } \left\|\sqrt{\frac{M_2}{M_1}}\alpha^T W_{y*}^{(d)} \right\|^2 =& 1  \mbox{ for }y=1,2,...,M_2, \label{eq:normw}
\end{eqnarray}
where the $d\times d'$ matrix $\alpha$ is the unknown. The bound in Eq.~(\ref{eq:SVbound}) is tight if and only if such matrix $\alpha$ solving this system of
equations can be found. Here $d$ is the degeneracy of the maximal singular value of $g$ and $d'$, the
dimension of the vectors $\vec{v}_x=\alpha^T V_{x*}^{(d)}$ and $\vec{w}_y=\alpha^T V_{y*}^{(d)}$, is a natural number. The steps leading to Eqs.~(\ref{eq:normv}) and (\ref{eq:normw}) can be found in the supplemental material of
\cite{PhysRevLett.111.240404}.
Because Eqs.~(\ref{eq:normv}) and (\ref{eq:normw}) are quadratic in $\alpha$ it may not be obvious
how to solve it. In \cite{PhysRevLett.111.240404} we described an algorithm to solve the above system of equations in polynomial time with
respect to the
size of $g$. The interested reader may also find a Matlab snippet in the Appendix~\ref{sec:snippets}. Often the solution $\alpha$ is
obvious, e.g. when it is proportional to the identity matrix.
\subsection{Optimal measurements are obtained from the SVD}\label{sec:measurements}
From the previous considerations we understand that the existence of the unit vectors $\vec{v}_x=\alpha^T V_{x*}^{(d)}$ and
$\vec{w}_y=\alpha^T V_{y*}^{(d)}$, i.e. the existence of the matrix $\alpha$ that allows this normalization, is crucial to the
satisfiability of the singular value bound. Furthermore they have a physical meaning,
because they are related to the observables in the following way.\\
Let us again consider the example of Eq.~(\ref{eq:CHSH}), with the singular value decomposition
\begin{equation}
 g=\left(\begin{array}{cc}
          1 & 1\\ 1 & -1
         \end{array}\right) = \underbrace{\left(\begin{array}{cc}
          \frac{1}{\sqrt{2}} & \frac{1}{\sqrt{2}}\\ \frac{1}{\sqrt{2}} & -\frac{1}{\sqrt{2}}
         \end{array}\right)}_{V} \underbrace{\left(\begin{array}{cc} 
\sqrt{2} & 0 \\
0 & \sqrt{2}
\end{array}\right)}_{S} \underbrace{\left(\begin{array}{cc} 
1 & 0 \\
0 & 1
\end{array}\right)}_{W^T}. \nonumber
\end{equation}
which we repeat from Eq.~(\ref{eq:SVDCHSH}). The
multiplicity $d=2$ of the maximal singular value $\sqrt{2}$ equals the number of measurement settings $M_1$ and $M_2$, so each of the rows
$V_{x*}^{(d)}$ and $W_{y*}^{(d)}$ are already normalized due to orthogonality of $V$ and $W$. Therefore we can choose
$\alpha=\mathds{1}^{(2)}$ to solve Eqs.~(\ref{eq:normv})
and
(\ref{eq:normw}). We then have $\vec{v}_1=(1,1)^T/\sqrt{2}$, $\vec{v}_2=(1,-1)^T/\sqrt{2}$, $\vec{w}_1=(1,0)^T$ and $\vec{w}_2=(0,1)^T$. We
are looking for a state and observables such that $E(x,y)=\vec{v}_x\cdot \vec{w}_y$, which is always possible to find (see Tsirelson's
theorem, Appendix~\ref{sec:proofofbound}).\\
Consider for example two spin-$\frac{1}{2}$ particles in the state $\vec{\phi}_+=\frac{1}{\sqrt{2}}(1,0,0,1)^T$. Alice and Bob can measure
their particles' spin with Stern-Gerlach apparatuses along any orientation in the $x$-$z$-plane. The observable of Alice corresponding to
a measurement along the direction $(a_x,a_z)^T$ is
\begin{equation}
 A=a_x \left(
\begin{array}{cc}
 0 & 1 \\
 1 & 0 \\
\end{array}
\right) + a_z \left(
\begin{array}{cc}
 1 & 0 \\
 0 & -1 \\
\end{array}
\right),
\end{equation}
where the matrices are two of the so-called Pauli matrices. Bob's measurement reads analogously. The reader can easily verify that the
expectation value of the joint observable $A\otimes B$ is given by
\begin{equation}
\vec{\phi}_+^\dagger (A\otimes B) \vec{\phi}_+ = \vec{a}\cdot \vec{b}.
\end{equation}
Therefore optimal measurement directions leading to equality in Ineq.~(\ref{eq:SVbound}) are given by $\vec{v}_x$ and $\vec{w}_y$. For this
reason we will call $\vec{v}_x$ and $\vec{w}_y$ the measurement directions, even though they can have a dimension greater than three for
general $g$.\\
We note how this construction of observables generalizes: The state can be taken to be $\vec{\phi}_+=\frac{1}{\sqrt{D}} \sum_{i=1}^D \vec{e}_i
\otimes \vec{e}_i$ and
the observables can be constructed as $A_x = \vec{v}_x \cdot \vec{X}$ and $B_y = \vec{v}_y \cdot \vec{X}$, where
$\vec{X}$ is a vector of matrices $X_i$ generalizing Pauli matrices in some sense (they anticommute, i.e. $X_i X_j + X_j X_i=0$ for $i\neq j$).
\subsection{Bell inequalities allow to lower bound the Hilbert space dimension} 
\begin{figure}
 \begin{center}
  \includegraphics[scale=0.6]{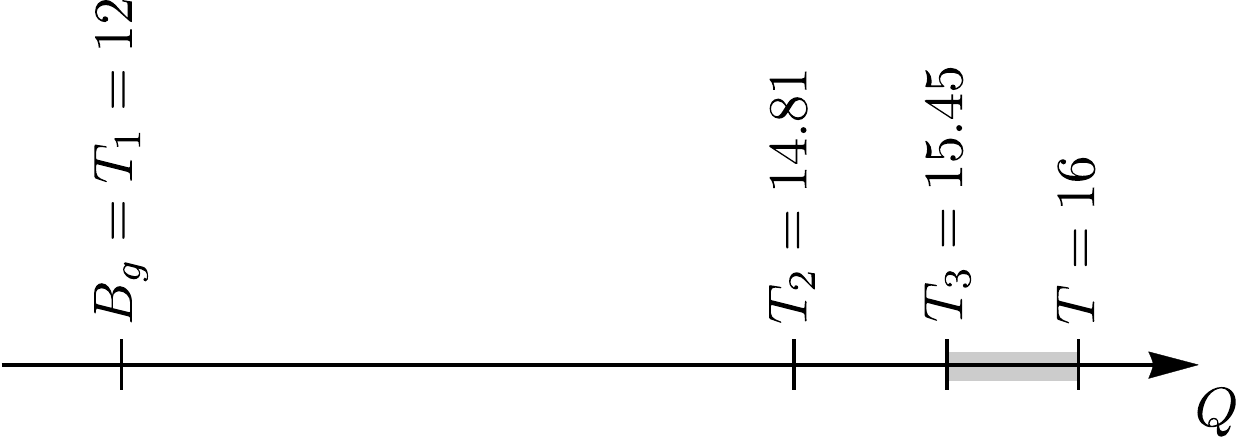}
 \end{center}
\caption{Depending on the dimension $d'$ of the measurement directions different values $T_{d'}$ are maximal for the Bell inequality given
by coefficients in Eq.~(\ref{eq:gBX4}). An experimentally obtained value $Q$ of the Bell inequality inside the shaded area witnesses,
that the produced quantum system had a greater Hilbert space dimension than qubits (see text). The values are taken from
\cite{PhysRevA.77.042106}.} \label{fig:allboundsBX4}
\end{figure}

In the previous example we chose $\alpha$ to be a square matrix, namely $\alpha=\mathds{1}^{(2)}$. We will now illustrate the role of the
dimension of the
measurement directions $d'$ with an example of a trivial Bell inequality, where $d'=1$ suffices to obtain the Tsirelson bound. For this
example the coefficients are $g=\mathds{1}^{(2)}$. An obvious singular value decomposition of this identity matrix is to choose
$V=S=W=\mathds{1}^{(2)}$. Just as before we can say that $\alpha=\mathds{1}^{(2)}$ is a solution to Eqs.~(\ref{eq:normv}) and
(\ref{eq:normw}), thus
the bound is achievable with $d'=2$. But we can also choose $\alpha=(1,1)^T$, which also solves the system of equations. In this case the
measurement directions are one-dimensional ($d'=1$), in fact they are all equal to $1$. Then the expectation value given by the scalar
product of the measurement directions reduces to the ``classical'' expectation value of deterministic local and realistic theories given in
Eq.~(\ref{eq:Eclass}). Both quantum theory and local realistic theories can achieve the maximal value of two. This inequality is therefore
unable to show a contradiction between quantum theory and locality, realism and freedom of choice. You might have expected this, since the
matrix of coefficients does not even contain a negative coefficient, which implies that the maximum value is achieved if all outcomes are
$+1$.\\
Let us discuss a more interesting example. It is a special instance of the family of Bell inequalities discussed by Vert\'esi and P\'al
in~\cite{PhysRevA.77.042106}. You can also find the following analysis for the whole family
in the supplemental material of \cite{PhysRevLett.111.240404}. The coefficients are
\begin{equation}
 g=\left(
\begin{array}{cccc}
 1 & 1 & 1 & 1 \\
 -1 & 1 & 1 & 1 \\
 1 & -1 & 1 & 1 \\
 -1 & -1 & 1 & 1 \\
 1 & 1 & -1 & 1 \\
 -1 & 1 & -1 & 1 \\
 1 & -1 & -1 & 1 \\
 -1 & -1 & -1 & 1 \\
\end{array}
\right).\label{eq:gBX4}
\end{equation}
Please note, that the columns of $g$ are orthogonal, thus it is easy to find a truncated singular value decomposition of $g$: We can
choose $V^{(d)}=\frac{1}{2\sqrt{2}} g$, $S^{(d)}=2\sqrt{2} \mathds{1}^{(4)}$ and $W=\mathds{1}^{(4)}$. One can easily check, that
$\alpha=\sqrt{2}\mathds{1}^{(4)}$ is a solution for the $(d\times d')$-matrix $\alpha$ of Eqs.~(\ref{eq:normv}) and (\ref{eq:normw}), so the maximal quantum value of $16$ (see Eq.~(\ref{eq:SVbound})) is
achievable with ($d'=4$)-dimensional measurement directions. It turns out, that the system of equations is not solvable if we choose $d'=3$, i.e. $\alpha$ to be a $(4\times 3)$-dimensional matrix. This has some very interesting physical implications. Since $(d'=3)$-dimensional measurement directions do not suffice to
obtain the maximal value of the Bell inequality, we can conclude from a measured value of $Q=16$, that our measurement directions were at
least four-dimensional. Of course one will never measure this value perfectly in experiment, so what one has to do in practice is to
calculate the maximum of the Bell inequality over all three-dimensional measurement directions (this is analog to the calculation of the
classical bound $B_g$ described above). If we call this value $T_3$, then any value between $T_3$ and $16$ witnesses the dimension of the
measurement directions to be at least four (see Figure~\ref{fig:allboundsBX4}).\\ 
For spin-$\frac{1}{2}$ particles, there are three orthogonal measurement directions (orientations of the Stern-Gerlach-apparatus), i.e.
$x$-, $y$- and $z$-direction, corresponding to the three Pauli matrices (see Eq.~(\ref{eq:obs})) and not more.
This holds for all quantum systems with two-dimensional Hilbert space (qubits). Thus if in some Bell experiment the value of the
Vert\'esi-P\'al-inequality given by the coefficients in Eq.~\ref{eq:gBX4} is found to be $16$ (or larger than $T_3$), one can conclude that
the produced and measured systems were no qubits. In particular they were not single spin-$\frac{1}{2}$ particles. Please note, that this
argument is independent of the physical implementation of the source and the measurement apparatuses. For this reason the concept is often
called device independent dimension witness.\\

\subsection{Satisfiability of the bound can be understood geometrically}
\begin{figure}[tbp]%
 \sidecaption[t]
\includegraphics[scale=0.6]{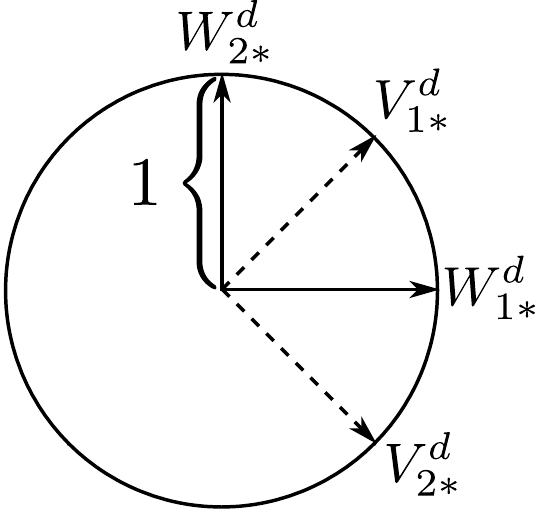} 
 \caption{The singular value bound is achievable if and only if the vectors $V_{x*}^{(d)}$ and $\sqrt{\frac{M_2}{M_1}} W_{y*}^{(d)}$ lie on
the
surface of an ellipsoid. These vectors and the ellipsoid (here a circle) are shown for the CHSH inequality.}\label{fig:kreise}%
\end{figure}

\begin{figure}[tbp]%
 \sidecaption
  \includegraphics[scale=0.6]{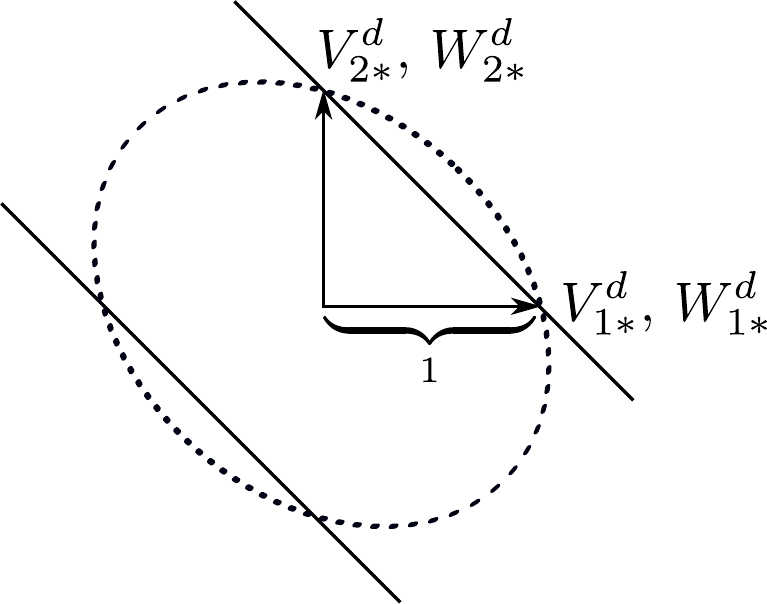} 
 \caption{The vectors $V_{x*}^{(d)}$ and $\sqrt{\frac{M_2}{M_1}} W_{y*}^{(d)}$ of $g=\mathds{1}^{(2)}$ lie on the dotted ellipse.
Increasing the larger semi-axis while keeping the vectors on the ellipse leads to the solid (degenerate) ellipse in the limit. Infinite
semi-axes of the ellipsoid imply, that lower dimensional measurement directions (here $d'=1$) suffice to achieve the Tsirelson
bound.}\label{fig:identitaet}%
\end{figure}

With $\vec{r}=V_{x*}^{(d)}$ Eq.~(\ref{eq:normv}) can be written as $\vec{r}^T \alpha \alpha^T \vec{r} = 1$. This quadratic form defines an
ellipsoid with semi-axes $\frac{1}{\sqrt{\lambda_1}},\frac{1}{\sqrt{\lambda_2}},...,\frac{1}{\sqrt{\lambda_d}}$ where
$\lambda_1,\lambda_2,...,\lambda_d$ are the eigenvalues of $\alpha \alpha^T$. Analogously the vectors $\vec{r}'=\sqrt{\frac{M_2}{M_1}}
W_{y*}^{(d)}$ lie on the same ellipsoid (see Eq.~(\ref{eq:normw})).\\
We therefore state, that the singular value bound is obtainable if and only if the vectors $V_{x*}^{(d)}$ and $\sqrt{\frac{M_2}{M_1}}
W_{y*}^{(d)}$
lie on an ellipsoid. As we mentioned before, in many cases (e.g. from the literature), $\alpha$ can be chosen to be proportional to
the identity matrix. Thus in these cases the vectors lie on a $d$-dimensional sphere, i.e. for $d=2$ they are on a circle, which is shown
for the CHSH inequality~\cite{PhysRevLett.23.880} in Fig.~\ref{fig:kreise}.\\
If $\alpha$ is not square or not full rank (i.e. at least one eigenvalue of $\alpha$ is zero), then at least one of the eigenvalues of $\alpha \alpha^T$ is zero, too. We define the corresponding
semi-axis to be infinite.\\
The measurement directions lie in the image of the linear transformation associated with $\alpha$. Thus the dimension of the measurement directions cannot be larger than the rank of
$\alpha$. For $g=\mathds{1}$ we show the degenerate ellipsoid with one infinite semi-axis corresponding to the solution $\alpha=(1,1)^T$
(see above) in Fig.~\ref{fig:identitaet}.
\subsection{Changing $g$ without changing the Tsirelson bound}
\begin{figure}[tbp]
 \begin{center}
  \includegraphics[scale=0.5]{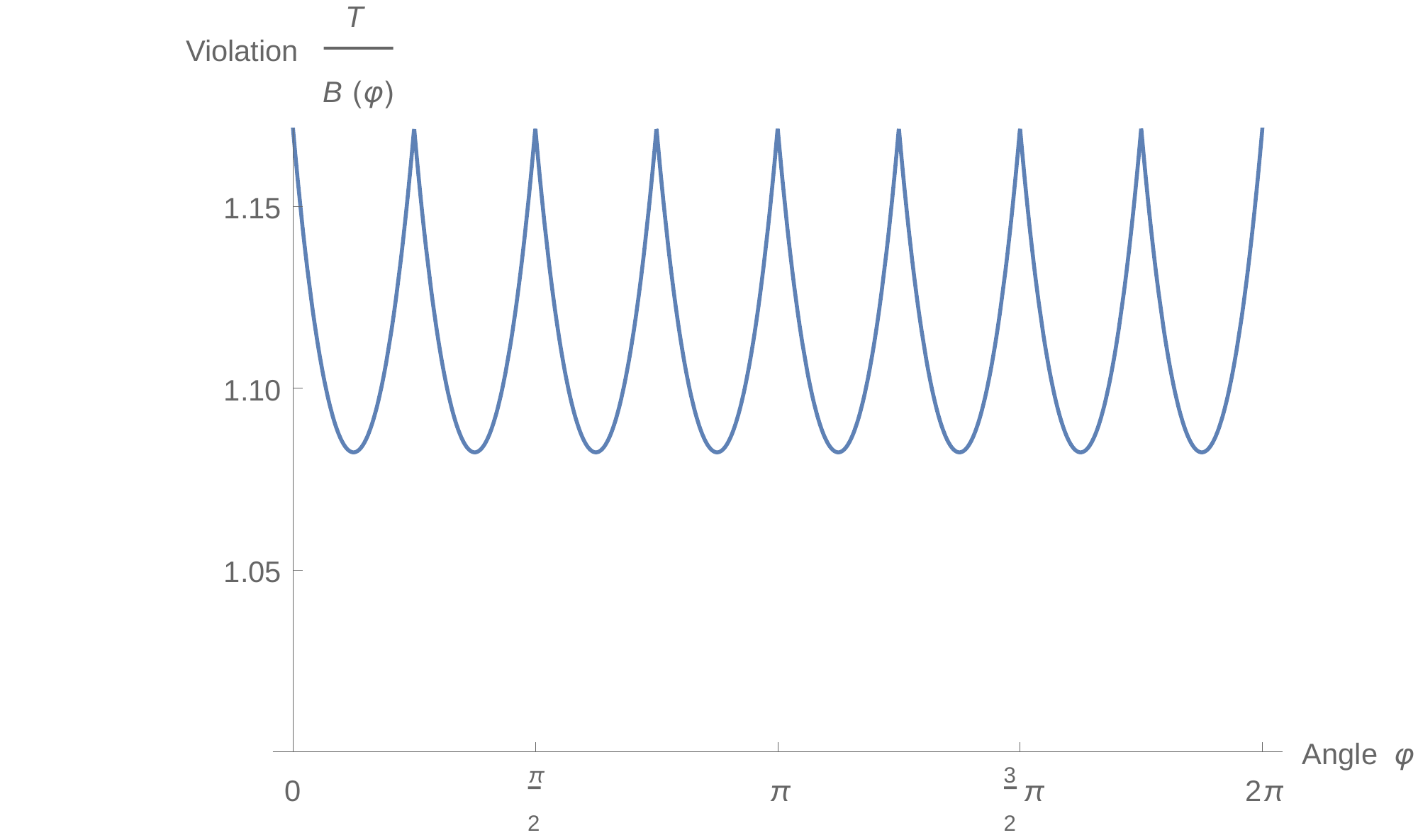}
 \end{center}
 \caption{The ratio of the maximal qauntum and classical value, the violation, is plotted for the Bell inequality
given by the coefficients of Eq.~(\ref{eq:gedrehteUngl}) as a function of the relative rotation of the two
laboratories $\varphi$.}\label{fig:gedrehteverletzung}
\end{figure}

The parts of the SVD of $g$ which do not correspond to the maximal singular value of (i.e. the non-shaded areas in Figure~\ref{fig:matrices}) did not appear in our discussion of the Tsirelson bound. Therefore
any changes of these singular vectors in $V$ and $W$ and singular values in $S$ will not affect our analysis. The last is, of course, only
true as long as these new singular values 
do not become bigger than the (previously) maximal singular value. While this changes the matrix $g$, i.e. leads to a new Bell
inequality, the quantum bound remains obtainable and its value remains the same.\\
From the geometric picture we immediately understand, that rotations of the vectors $V_{x*}^{(d)}$ and $\sqrt{\frac{M_2}{M_1}} W_{y*}^{(d)}$
which
keep them on the ellipsoid (see Figs. \ref{fig:kreise} and \ref{fig:identitaet}) also do not change the value and satisfiability of the singular value bound.\\
\begin{figure}[tbp]
 \begin{center}
  \includegraphics[width=0.7\textwidth]{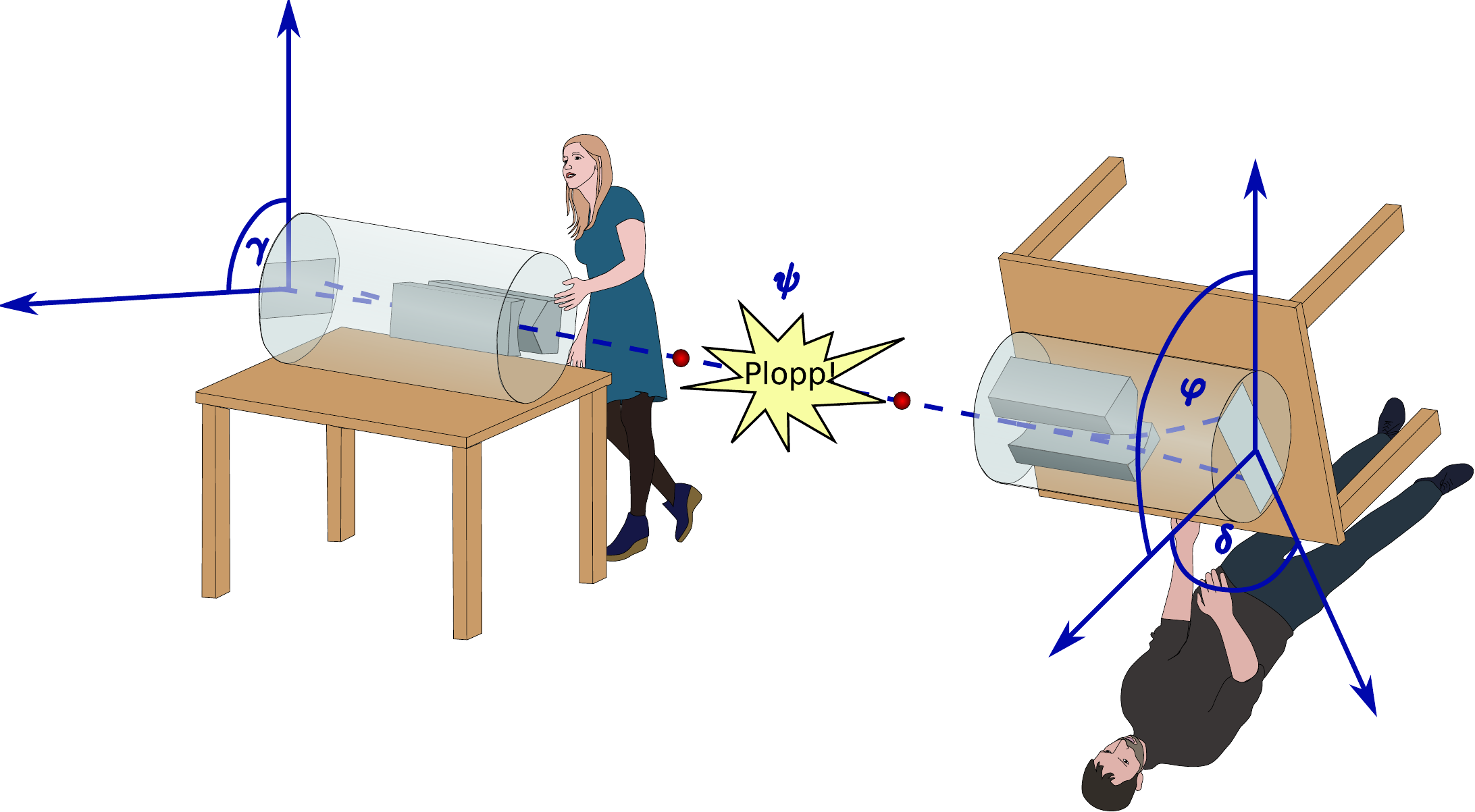}
 \end{center}
 \caption{Alice and Bob share pairs of particles in a spin-entangled state $\psi$ and want to violate a Bell inequality. They
each can measure the spin of their particle along transversal axes with different angle relative to the table's up.
Unfortunately they were not able to agree on what ``up'' means, yet, and their local coordinate systems are twisted by a relative angle
$\varphi$. The text explains that one possibility is to measure using (local) angles $\gamma_1=45^\circ$, $\gamma_2=-45^\circ$,
$\gamma_3=0^\circ$, $\gamma_4=90^\circ$ at Alice's site and $\delta_1=0^\circ$ and $\delta_2=90^\circ$ at Bob's site and ``rotate'' the Bell
inequality.}\label{fig:gedrehtetische}
\end{figure}
\noindent We give an example to illustrate that the measurement directions can be rotated without affecting the singular value bound and its
tightness. Consider the CHSH test described above, but now Alice and Bob did not agree on a common coordinate system before
performing the experiment, see Figure~\ref{fig:gedrehtetische}. Let us assume for simplicity that their local coordinate systems are only
rotated relative to each other by an angle $\varphi$ around their common $y$-axis. This angle $\varphi$ is unknown to Alice and Bob at the
time of collecting the measurement data. The quantum state is still $\vec{\psi}=\frac{1}{\sqrt{2}}(1,0,0,1)^T$, independent of $\varphi$.\\
Let us analyze the effect of the relative rotation on the violation of the CHSH inequality. The first idea might be to measure the
observables of Section~\ref{sec:measurements} in the local basis and insert the estimated expectation values into the CHSH inequality. For a
relative angle
$\varphi=0^\circ$ these observables are optimal, but for an angle of $\varphi=-45^\circ$ Alice and Bob measure in the same direction and
their data will not violate the CHSH inequality. From the previous considerations we know that it is also possible to ``rotate'' the Bell
inequality such that the actually performed measurements are optimal for that inequality. This is can be done by applying a rotation matrix
to the matrix $W$. However, twisting the original CHSH inequality by
$45^\circ$ gives $\sqrt{2}\mathds{1}$ (up to relabeling of the measurement settings), see Figures~\ref{fig:kreise} and \ref{fig:identitaet}.
And as it is shown in Figure~\ref{fig:identitaet} all but one semiaxis of the ellipse associated with $\alpha$ can be chosen to be
infinite, which is equivalent to the fact that the classical bound and the quantum bound coincide. This implies that the inequality given by
coefficients $g=\sqrt{2}\mathds{1}$ cannot be violated.\\
The trick is to include more measurement directions. If the measurement directions of Alice already uniquely define the ellipsoid associated
with $\alpha$, then the rotation of the measurement directions of Bob does not change the fact that the Bell inequality can be violated.
One obvious possibility to achieve this is to add all settings of Bob to Alice. We do this for the CHSH inequality (see
Eq.~(\ref{eq:SVDCHSH})) and get
\begin{equation}
 g(\varphi)=\left(\begin{array}{cc}
                   \frac{1}{\sqrt{2}} & \frac{1}{\sqrt{2}}\\
                   \frac{1}{\sqrt{2}} & -\frac{1}{\sqrt{2}}\\
                   1 & 0\\
                   0 & 1
                  \end{array}\right) \left(\begin{array}{cc}
\cos(\varphi) & -\sin(\varphi)\\
\sin(\varphi) & \cos(\varphi)
\end{array}
\right). \label{eq:gedrehteUngl}
\end{equation}
If we call the different measurement angles $\gamma_1,\gamma_2,\gamma_3,\gamma_4$ at Alice's site and $\delta_1,\delta_2$ at
Bob's site we have for $\alpha=\sqrt{2}\mathds{1}$ that $\gamma_1=45^\circ$, $\gamma_2=-45^\circ$, $\gamma_3=0^\circ$,
$\gamma_4=90^\circ$, $\delta_1=0^\circ$ and $\delta_2=90^\circ$ are optimal measurement settings. The quantum value $T=4$ of this inequality
does not depend on $\varphi$, but the classical bound $B$ does. Figure~\ref{fig:gedrehteverletzung} shows the violation of the Bell
inequality depending on the relative rotation $\varphi$. As expected it is always strictly larger than one. The maximal violation of $4 - 2
\sqrt{2}$ can be obtained for $\varphi=k \frac{\pi}{4}$, where $k$ is an integer number. We remark that if Alice and Bob even do not agree
on a common coordinate system for the analysis of the data, they still can maximize the violation over the angle $\varphi$.\\
A similar analysis for a general rotation in three dimensions given by three Euler angles was done in~\cite{ModifiedBI}.
Different approaches to Bell inequalities without a common coordinate system have been described in the literature. We want to mention the
following strategy. Each party measures along random but orthogonal measurement directions. Afterwards the violation of the CHSH inequality
is calculated for all combinations of pairs of measured settings of Alice and Bob. The result is similar to the one in this section: if the
parties measure along more than two directions, then one can find a Bell inequality that is violated with
certainty~\cite{GuaranteedViolation}.\\
A deeper understanding of the correlations between measurements on separated systems possible according to quantum theory, including the maximal value of Bell inequalities, is an aim of ongoing research in the field of quantum information theory. In this essay we saw how more measurement settings and higher-dimensional quantum systems can lead to stronger violations of Bell inequalities, e.g. in the context of device-independent dimension witnesses or Bell experiments without a shared reference frame. The insights gained from these simple examples may help to find Bell inequalities well suited for different situations and applications.
\begin{acknowledgement}
We thank Jochen Szangolies and Michaela Stötzel for feedback which helped to improve this manuscript. ME acknowledges financial support of BMBF, network Q.com-Q. 
\end{acknowledgement}

\bibliographystyle{spphys} 

\begin{thebibliography}{10}
\providecommand{\url}[1]{{#1}}
\providecommand{\urlprefix}{URL }
\expandafter\ifx\csname urlstyle\endcsname\relax
  \providecommand{\doi}[1]{DOI \discretionary{}{}{}#1}\else
  \providecommand{\doi}{DOI \discretionary{}{}{}\begingroup
  \urlstyle{rm}\Url}\fi

\bibitem{YoungLectures}
T.~Young, \emph{A Course of Lectures on Natural Philosophy and the Mechanical
  Arts} (Taylor and Walton, 1845)

\bibitem{FeynmanLectures3}
R.P. Feynman, R.B. Leighton, M.~Sands, \emph{The Feynman Lectures on Physics},
  vol.~3 (Addison Wesley, 1971).
\newblock \urlprefix\url{http://www.feynmanlectures.caltech.edu/}

\bibitem{NJP153033018}
R.~Bach, D.~Pope, S.H. Liou, H.~Batelaan, New Journal of Physics
  \textbf{15}(3), 033018 (2013).
\newblock \doi{10.1088/1367-2630/15/3/033018}

\bibitem{PhysRevLett.88.100404}
B.~Brezger, L.~Hackerm\"uller, S.~Uttenthaler, J.~Petschinka, M.~Arndt,
  A.~Zeilinger, Phys. Rev. Lett. \textbf{88}, 100404 (2002).
\newblock \doi{10.1103/PhysRevLett.88.100404}

\bibitem{NaturePhysics.10.271}
M.~Arndt, K.~Hornberger, Nature Physics \textbf{10}, 271 (2014).
\newblock \doi{10.1038/nphys2863}

\bibitem{Razavy}
M.~Razavy, \emph{Quantum Theory of Tunneling} (World Scientific Pub Co Inc,
  2003)

\bibitem{FeynmanLectures2}
R.P. Feynman, R.B. Leighton, M.~Sands, \emph{The Feynman Lectures on Physics},
  vol.~2 (Addison Wesley, 1977).
\newblock \urlprefix\url{http://www.feynmanlectures.caltech.edu/}

\bibitem{PhysRevA.41.2295}
W.M. Itano, D.J. Heinzen, J.J. Bollinger, D.J. Wineland, Phys. Rev. A
  \textbf{41}, 2295 (1990).
\newblock \doi{10.1103/PhysRevA.41.2295}

\bibitem{BombTester}
A.C. Elitzur, L.~Vaidman, Foundations of Physics \textbf{23}(7), 987 (1993).
\newblock \doi{10.1007/BF00736012}

\bibitem{PhysRevLett.83.4725}
P.G. Kwiat, A.G. White, J.R. Mitchell, O.~Nairz, G.~Weihs, H.~Weinfurter,
  A.~Zeilinger, Phys. Rev. Lett. \textbf{83}, 4725 (1999).
\newblock \doi{10.1103/PhysRevLett.83.4725}

\bibitem{QuantumImaging}
G.~Barreto~Lemos, V.~Borish, G.D. Cole, S.~Ramelow, R.~Lapkiewicz,
  A.~Zeilinger, Nature pp. 409--412 (2014)

\bibitem{PhysRevLett.49.91}
A.~Aspect, P.~Grangier, G.~Roger, Phys. Rev. Lett. \textbf{49}, 91 (1982).
\newblock \doi{10.1103/PhysRevLett.49.91}

\bibitem{PhysRevLett.81.5039}
G.~Weihs, T.~Jennewein, C.~Simon, H.~Weinfurter, A.~Zeilinger, Phys. Rev. Lett.
  \textbf{81}, 5039 (1998).
\newblock \doi{10.1103/PhysRevLett.81.5039}

\bibitem{Wineland}
M.A. Rowel, D.~Kielpinski, V.~Meyer, C.A. Sackett, W.M. Itano, C.~Monroe, D.J.
  Wineland, Nature \textbf{409}, 791 (2001).
\newblock \doi{10.1038/35057215}

\bibitem{Nature461.504}
J.M.{\it et. al.}. Martinis, Nature \textbf{461}, 504 (2009).
\newblock \doi{10.1038/nature08363}

\bibitem{JPA46424009}
M.~Zukowski, C.~Brukner, Journal of Physics A: Mathematical and Theoretical
  \textbf{47}(42), 424009 (2014).
\newblock \doi{10.1088/1751-8113/47/42/424009}

\bibitem{PhysRevLett.23.880}
J.F. Clauser, M.A. Horne, A.~Shimony, R.A. Holt, Phys. Rev. Lett. \textbf{23},
  880 (1969).
\newblock \doi{10.1103/PhysRevLett.23.880}

\bibitem{PhysRevD.10.526}
J.F. Clauser, M.A. Horne, Phys. Rev. D \textbf{10}, 526 (1974).
\newblock \doi{10.1103/PhysRevD.10.526}

\bibitem{GerthsenPhysik}
C.~Gerthsen, \emph{Physik} (D. Meschede, 2001)

\bibitem{UndergraduateLaboratory}
D.~Dehlinger, M.W. Mitchell, Am. J. Phys. \textbf{70}, 903 (2002).
\newblock \doi{10.1119/1.1498860}

\bibitem{Cirel'son1980}
B.~Tsirelson, Lett. Math. Phys. \textbf{4}, 93 (1980)

\bibitem{Wehner}
S.~Wehner, Phys. Rev. A \textbf{73}, 022110 (2006).
\newblock \doi{10.1103/PhysRevA.73.022110}

\bibitem{PhysRevLett.98.010401}
M.~Navascu\'es, S.~Pironio, A.~Ac\'in, Phys. Rev. Lett. \textbf{98}, 010401
  (2007).
\newblock \doi{10.1103/PhysRevLett.98.010401}

\bibitem{PopescuRohrlich}
S.~Popescu, D.~Rohrlich, Foundations of Physics \textbf{24}(3), 379 (1994).
\newblock \doi{10.1007/BF02058098}

\bibitem{InformationCausality}
M.~Pawlowski, T.~Paterek, D.~Kaslikowski, V.~Scarani, A.~Winter, M.~Zukowski,
  Nature pp. 1101--1104 (2009).
\newblock \doi{10.1038/nature08400}

\bibitem{PhysRevLett.110.060402}
A.~Cabello, Phys. Rev. Lett. \textbf{110}, 060402 (2013).
\newblock \doi{10.1103/PhysRevLett.110.060402}

\bibitem{PhysRevLett.111.240404}
M.~Epping, H.~Kampermann, D.~Bru{\ss}, Phys. Rev. Lett. \textbf{111}, 240404
  (2013).
\newblock \doi{10.1103/PhysRevLett.111.240404}

\bibitem{PhysRevA.77.042106}
T.~V\'ertesi, K.F. P\'al, Phys. Rev. A \textbf{77}, 042106 (2008).
\newblock \doi{10.1103/PhysRevA.77.042106}

\bibitem{ModifiedBI}
M.~Epping, H.~Kampermann, D.~Bru{\ss}, J. Phys. A: Math. Theor.
  \textbf{47}(424015) (2014).
\newblock \doi{10.1088/1751-8113/47/42/424015}

\bibitem{GuaranteedViolation}
P.~Shadbolt, T.~V\'ertesi, Y.C. Liang, C.~Branciard, N.~Brunner, J.L. O'Brien,
  Scientific Reports \textbf{2}(470) (2012).
\newblock \doi{10.1038/srep00470}

\bibitem{Tsirelson1980}
B.~Tsirelson, Letters in Mathematical Physics pp. 93--100 (1980)

\end{thebibliography}

\section*{Appendix}
\addcontentsline{toc}{section}{Appendix}
\section{Tsirelson's theorem carries the Tsirelson bound to Linear Algebra}\label{sec:proofofbound}
We now sketch the derivation of Eq.~(\ref{eq:SVbound}) following \cite{PhysRevLett.111.240404}. It is strongly based on a theorem by Boris
Tsirelson \cite{Tsirelson1980}. It links the expectation values of quantum measurements to scalar products of real
vectors. While the full theorem shows
equivalence of five different ways of expressing the expectation value, we will repeat two of them here.\\
Remember that in the formalism of quantum theory observables are hermitean operators, i.e. they equal their complex conjugated transpose.
And quantum states can be described by density matrices, which are convex mixtures of projectors onto pure quantum states, with the weights
being
the probability to find the system in the corresponding pure state. This implies that the density matrix is positive and has trace one.\\
Consider two fixed sets of observables with eigenvalues in $[-1,1]$, $\{A_1,A_2,...,A_{M_1}\}$ and
$\{B_1,B_2,...,B_{M_2}\}$, and a quantum state given in terms of its density matrix $\rho$. Then the
expectation value of the joint measurement of $A_x$ and $B_y$, $A_x\otimes B_y$,  is 
$E(x,y)=\mathrm{tr} (A_x \otimes B_y \rho)$ according to quantum theory. Tsirelson's theorem states, that there exist
real $M_1+M_2$ dimensional unit vectors $\{\vec{v}_1,\vec{v}_2,...,\vec{v}_{M_1}\}$ and $\{\vec{w}_1,\vec{w}_2,...,\vec{w}_{M_2}\}$ such
that all expectation values can be expressed as $E(x,y)=\vec{v}_x \cdot \vec{w}_y$. This is the direction we need, because it allows us to
replace the expectation value in Eq.~(\ref{eq:bellineq}) by the scalar product of some real vectors. Tsirelson also proved the converse
direction: given the vectors $\vec{v}_1,\vec{v}_2,...,\vec{v}_{M_1}$ and $\vec{w}_1,\vec{w}_2,...,\vec{w}_{M_2}$ there exist observables
$A_1,A_2,...,A_{M_1}$ and $B_1,B_2,...,B_{M_2}$ and a state $\rho$ such that the
expectation value $E(x,y)=\mathrm{tr} (A_x \otimes B_y \rho)$ equals the scalar product $\vec{v}_x \cdot \vec{w}_y$.
\\
After application of Tsirelson's theorem Eq.~(\ref{eq:bellineq}), i.e. $\sum_{x,y} g_{x,y} E(x,y)$, takes the form
\begin{eqnarray}
 \sum_{x=1}^{M_1} \sum_{y=1}^{M_2} g_{x,y} \sum_{i=1}^{M_1+M_2} v_{x,i} w_{y,i} &=& \sum_{x=1}^{M_1} \sum_{y=1}^{M_2}
\sum_{i=1}^{M_1+M_2} \sum_{j=1}^{M_1+M_2}  v_{x,i}     g_{x,y} \delta_{ij}         w_{y,j}\nonumber\\
&=& \vec{v}^T (g\otimes \mathds{1}^{(M_1+M_2)}) \vec{w}. \label{eq:withtsirelsonvectors}
\end{eqnarray}
Here we expressed the scalar product as a matrix product using the $M_1+M_2$ dimensional identity matrix $\mathds{1}^{(M_1+M_2)}$ and
defined
the vectors $\vec{v}$ and $\vec{w}$, which arise if one concatenates all $\vec{v}_x$ and $\vec{w}_y$, respectively.
For the decomposition given in Eq.~(\ref{eq:SVDCHSH}) with $\alpha=\mathds{1}^{(2)}$, for example,
$\vec{v}_1=(\frac{1}{\sqrt{2}},\frac{1}{\sqrt{2}})^T$ and
$\vec{v}_2=(\frac{1}{\sqrt{2}},-\frac{1}{\sqrt{2}})^T$ and
thus $\vec{v}=(\frac{1}{\sqrt{2}},\frac{1}{\sqrt{2}},\frac{1}{\sqrt{2}},-\frac{1}{\sqrt{2}})^T$.
From Eq.~(\ref{eq:withtsirelsonvectors}) we see, that the maximal quantum value of the Bell inequality is given by the maximal singular
value (the maximal stretching factor) of $g\otimes \mathds{1}^{(M_1+M_2)}$ times the length of the vectors $\vec{v}$ and $\vec{w}$. The
matrix $g\otimes \mathds{1}^{(M_1+M_2)}$ has the same singular values as $g$, except that each of them appears $M_1+M_2$ times. Because the
$\vec{v}_x$ and $\vec{w}_y$ constituting $\vec{v}$ and $\vec{w}$ are all unit vectors, the length of $\vec{v}$ is $\sqrt{M_1}$ and the
length of $\vec{w}$ is $\sqrt{M_2}$. Putting these factors together we arrive at Eq.~(\ref{eq:SVbound}).\\
\section{MATLAB snippets}\label{sec:snippets}

\begin{lstlisting}[language=Matlab]
function [ T ] = singularvaluebound( g )
%SINGULARVALUEBOUND Calculates the SV-bound of g
%   the returned value is a Tsirelson bound for 
%   the CHSH-type inequality given by g
T=sqrt(numel(g))*norm(g);
end

function [ a ] = alphamatrix( g )
%ALPHAMATRIX Links SVD to measurement directions
%   see PRL 111, 240404 (2013)
[M1 M2]=size(g);
[V S W]=svd(g);
acc=1E-4; % adjust to numerical precision
d=sum(diag(S)>=S(1,1)-acc); 
% the vectors to be normalized by alpha:
A=[V(1:M1,1:d); sqrt(M2/M1)*W(1:M2,1:d)];
Q=(A*A').^2; 
c=pinv(Q)*ones(M1+M2,1);
if sum(abs(Q*c-ones(M1+M2,1))>acc)
    error('alphamatrix:nosol','No solution alpha found.');
else
    X=A'*diag(c)*A;
    if eigs(X,1,'sm')<0
        error('alphamatrix:norealsol',
              'No real solution alpha found.');
    end
end
a=X^0.5;
end

function [ T ] = tsirelsonbound( g )
%TSIRELSONBOUND Calculates the Tsirelson bound for g
%   Uses the semidefinite programm described by 
%   Stephanie Wehner in PRA 73, 022110 (2006).
[M1 M2]=size(g);
W=[zeros(M1,M1) g; g' zeros(M2,M2)];
G=sdpvar(M1+M2,M1+M2);
obj=trace(G*W)/2;
F=set(G>0);
for i=1:M1+M2
    F=F+set(G(i,i) == 1);
end
solvesdp(F,obj,sdpsettings('verbose',0));
T=-double(obj); 
end
\end{lstlisting}
\end{document}